\documentclass[aps,pra,twocolumn,showpacs,groupedaddress]{revtex4-1}

\usepackage[T1]{fontenc}
\usepackage{graphicx}
\usepackage{color,ulem}

\newcommand\forget[1]{}

\newcommand\ket[1]{|#1\rangle}

\begin{document}


\title{A quantum-enabled Rydberg atom electrometer}


\author{Adrien Facon}
\author{Eva-Katharina Dietsche}
\author{Dorian Grosso}
\author{Serge Haroche}
\author{Jean-Michel Raimond}
\author{Michel Brune}
\author{S\'ebastien Gleyzes}
\affiliation{Laboratoire Kastler Brossel, Coll\`ege de France, CNRS, ENS-PSL Research
University, UPMC-Sorbonne Universit\'es, 11, place Marcelin Berthelot, 75231 Paris Cedex 05, France}

\date{\today}


\maketitle

{\bf
There is no fundamental limit to the precision of a classical measurement. The position of a meter's needle can be determined with an arbitrarily small uncertainty. In the quantum realm, however, fundamental quantum fluctuations due to the Heisenberg principle limit the measurement precision. The simplest measurement procedures, involving semi-classical states of the meter, lead to a fluctuation-limited imprecision at the standard quantum limit \cite{Itano1993,Giovannetti2011}. By engineering the quantum state of the meter system, the measurement imprecision can be reduced down to the fundamental Heisenberg Limit (HL). Quantum-enabled metrology techniques are thus in high demand and the focus of an intense activity \cite{Wasilewski2010,Leibfried2004,Nagata2007,Jones2009,Mussel2014,Bohnet2014,Lo2015,Barontini2015,Tanaka2015,Hosten2016}. We report here a quantum-enabled measurement of an electric field based on this approach. We cast Rydberg atoms in Schr\"odinger cat states, superpositions of atomic levels with radically different polarizabilites. We use a quantum interference process to perform a measurement close to the HL \cite{Giovannetti2011}, reaching a single-shot sensitivity of 1.2~mV/cm for a 100~ns interaction time, corresponding to 30~$\mu$V/cm/$\sqrt{\mbox{Hz}}$ at our 3 kHz repetition rate. This highly sensitive, non-invasive space- and time-resolved field measurement extends the realm of electrometric techniques \cite{Dolde2011,Vamivakas2011,Houel2012,Dolde2014,Arnold2014} and could have important practical applications. Detection of individual electrons in mesoscopic devices \cite{Cleland1998,Bunch2007,Yoo1997, Devoret2000} at a $\simeq100\ \mu$m distance, with a MegaHertz bandwith is within reach.
}

Quantum metrology aims at measuring a classical quantity $A$ (a frequency, a field...) with the highest precision compatible with the quantum limits. It make use of a meter system, whose evolution depends upon $A$. This meter is initially prepared in a reference state and, after some interrogation time $\tau$, read out by a projective measurement. The standard approach uses as a meter an ensemble of $N$ two-level atoms (or spin-$1/2$ systems) evolving independently. Each of them undergoes, for instance, a Ramsey interferometric sequence. After $k$ repetitions of the experiment, $A$ is determined with a precision at the Standard Quantum Limit (SQL), scaling as  $1/\sqrt{Nk}$ \cite{Itano1993}.

The precision can be enhanced beyond the SQL by entangling the $N$ spins \cite{Giovannetti2011}. In most practical situations, the $N$ spin-$1/2$ systems are equivalent and their symmetric states can be described as those of a large $J=N/2$ spin \cite{Arecchi1972}. Using spin squeezed states, the HL, scaling as $1/(N\sqrt k)$, can then be approached \cite{Mussel2014,Bohnet2014,Hosten2016}. It can even be reached with Schr\"odinger cat-like states such as $(|\uparrow\rangle^{\otimes N}+|\downarrow\rangle^{\otimes N})$. However, the preparation of cat states \cite{Massar2003,Lau2014} is experimentally challenging \cite{Monz2011,Signoles2014}. Their practical metrological use \cite{Tanaka2015} has been restricted so far to $N\sim 10$ particles \cite{Leibfried2004, Nagata2007,Jones2009}.

Our alternative strategy uses directly a meter system made of a large spin $J$ carried by a single atom. When its evolution is quasi-classical, proceeding through Spin Coherent States (SCS) \cite{Arecchi1972}, the measurement precision is limited by the SQL, scaling as $1/\sqrt{2Jk}$. The HL for this large spin meter, however, scales as $1/(2J\sqrt k)$ \cite{Giovannetti2011}. It can be approached when the spin undergoes a non-classical evolution, for instance through Schr\"odinger cat states.

We report here a quantum-enabled measurement based on this principle. It determines the amplitude $F$ of an electric field $\mathbf F$ oriented along the $Oz$ quantization axis. The spin-$J$ system belongs to the Rydberg manifold $n=50$ of a rubidium atom. Rydberg atoms have a very large polarizability, which makes them particularly suitable for measurements of small electric fields \cite{Osterwalder1999, Abel2011}. The Stark levels in the manifold can be sorted by their magnetic quantum number $m$ [Fig. 1 (a)]. We use the ladder made up of the lowest energy levels for each $m$, equidistant to first order in $F$. This ladder is equivalent to that of the $|J,M\rangle$ levels of a spin $J=(n-1)/2=49/2$ with $M=m-J$. This spin evolves on a generalized Bloch sphere $\cal B$, $|J,J\rangle$ being the circular Rydberg state $nC$ at the north pole of $\cal B$.  The spin levels are connected by $\sigma_+$-polarized radio-frequency (rf) transitions at the frequency $\omega(F)/2\pi=3nFea_0/2h$ ($a_0$: Bohr radius), with $\partial\omega(F)/\partial F=2\pi\cdot 96$~MHz/(V/cm). The spin coherent states \cite{Arecchi1972}, $\ket{\theta,\varphi}$, corresponding to a Bloch vector pointing in the $\theta,\varphi$ direction on $\cal B$ are defined by  $\ket{\theta,\varphi}={\cal R}(\theta,\varphi)\ket{J,J}$. The rotation operator ${\cal R}(\theta,\varphi)$ is realized by the application of a nearly resonant classical rf field with a Rabi frequency $\Omega_{rf}$ and a phase $\varphi$ for a duration $t$ such that $\theta=\Omega_{rf} t$.

A measurement at the SQL using SCS relies on a double rf pulse technique (Ramsey scheme, Fig 1.b). A first rf pulse at  $\omega_{rf}$ prepares the $|\theta,0\rangle$ SCS from the initial state $|J,J\rangle$. In a frame rotating at $\omega_{rf}$ around $Oz$, the further spin evolution is a precession at $\omega(F)-\omega_{rf}$, leading after an interrogation time $\tau$ to the $\ket{\theta,\phi}$ SCS with $\phi= (\omega(F)-\omega_{rf}) \tau$. The field-sensitive phase $\phi$ is then read out by applying a final rotation $\mathcal R(\theta,\pi+\varphi_{rf})$ with an adjustable phase $\varphi_{rf}$. A measurement of the final spin state $\ket{\psi_f}$ provides information on $F$ with a variance $\sigma_{F,SQL}=\sigma_{F,SQL}^1/\sqrt k$, where $\sigma_{F,SQL}^1$ is the single-shot SQL sensitiviy \cite{Itano1993}:
\begin{equation}
\sigma_{F,SQL}^1=\frac{1}{\tau\sqrt{2J}}\left(\frac{\partial\omega}{\partial F}  \right)^{-1}\ .
\label{eqn-sql}
\end{equation}

In order to beat the SQL, we measure, instead of $\phi$, the global quantum phase $\Phi$ accumulated by the spin $J$ during its evolution on $\cal B$. Measuring $\Phi$ as a function of $F$ requires a quantum reference state $|R\rangle$, unaffected by the spin $J$ successive transformations. We use, for $|R\rangle$, the $n=51$ circular state (Fig. 1.a). 
We initially prepare the superposition $1/\sqrt 2 (|J,J\rangle+|R\rangle)$ using a classical microwave (mw) pulse. The $|J,J\rangle$ part of this initial state then undergoes the Ramsey sequence sketched in Fig 1.b, ending in state $\ket{\psi_f}$, $\Phi$ being the phase of $\langle \psi_f \ket{J,J}$. We then apply a second $\pi/2$ mw pulse, selectively addressing the $|J,J\rangle\rightarrow\ket R$ transition, with an adjustable phase $\varphi_{mw}$. We finally measure whether the atom is in the state $|J,J\rangle$ or not (Methods). 

The probability for finding the atom in $|J,J\rangle$ oscillates as a function of $\Phi$. This oscillation is sensitive to small variations of $F$, since the atomic system is cast during the interrogation time $\tau$ in a quantum superposition of two states with different static dipoles, $|R\rangle$ and $\ket{\theta,\varphi}$, an atomic cat state \cite{Hempel2013}. The interference phase $\Phi$ depends upon the exact spin trajectory on $\cal B$ and thus upon $F$ and $\varphi_{rf}$ (Methods). The amplitude of the interference pattern is proportional to $|\langle \psi_f \ket{J,J}|$. It is maximum for $\varphi_{rf}\approx 0$ when the field $F$ is close to the reference field $F_0$ such that $\omega(F_0)=\omega_{rf}$. Then, $\Phi$ can be expanded to first order in a small field variation $dF=F-F_0$ as:
\begin{equation}
\Phi \approx \Phi_0 + J \,(1 - \cos\theta)\,\left(\frac{\partial\omega}{\partial F}  \right) dF \,\,  \tau\ \ ,
\label{eqn-phicat}
\end{equation}
where $\Phi_0$ is the total phase accumulated for the reference field $F_0$ (Methods). This leads to a single-shot measurement sensitivity:
\begin{equation}
\sigma_{F}^1= \frac{1}{\tau J(1 - \cos\theta)}\left(\frac{\partial\omega}{\partial F}  \right)^{-1}\ ,
\end{equation}
scaling as $1/J$. The factor $J(1-\cos\theta)$, proportional to the difference of the electric dipoles of the  components of the superposition, measures the `size' of the Schr\"odinger cat. The Heisenberg limit, $\sigma_{F,HL}^1= (1/2J\tau)(\partial \omega/\partial F)^{-1}$, is reached for $\theta = \pi$, when this size is maximum.

In the real experiment, we must take into account the finite duration of the rf pulses ($\Omega_{rf}/2\pi = 1.6$~MHz) and the second order Stark effect in the $n=50$ manifold, which makes the spin states ladder slightly anharmonic. The trajectory of the spin on $\cal B$ and the spin coherent states are distorted accordingly (Fig. 1.c). The optimal phase for the second rf pulse is thus $\varphi_{rf}=\varphi_{rf}^0\not=0$. The state distortion slightly affects the contrast of the interferometric signal. Nevertheless, the main conclusions of the simple case discussion above remain valid.

Fig. 2(a) presents, for reference, the results of the classical Ramsey method, in which no microwave pulses are applied (timing in the inset). We measure the probability $P(\varphi_{rf})$ for returning in $|J,J\rangle$ as a function of $\varphi_{rf}$ for two electric fields $F_0-\delta F/2$ and $F_0+\delta F/2$, with $\delta F=566\ \mu$V/cm. These probabilities are Gaussian (Methods) centered around $\varphi_{rf}^0=0.715$~rd. The contrast is slightly reduced and the width increased w.r.t the ideal case due to the second order Stark effect. The phase shift ($\delta\phi=82$~mrad) induced by the variation $\delta F$ of the electric field is small as compared to the width of the signal ($\approx 0.722$~rd). 

Let us now consider the complete sequence [timing in the inset of Fig. 2(b)], with a fixed mw phase $\varphi_{mw}=\varphi_{mw}^0$. We measure the probability $P(\varphi_{rf},\varphi_{mw}^0)$ to detect finally the atom in the initial $|J,J\rangle$ state as a function of $\varphi_{rf}$ for the electric fields $F_0-\delta F/2$ and $F_0+\delta F/2$. This probability exhibits an interference pattern around $\varphi_{rf}=\varphi_{rf}^0$, revealing the rapid variation of $\Phi$ with $\varphi_{rf}$ (Methods). The contrast of the interference reflects the probability amplitude for the spin $J$ to return in its initial $|J,J\rangle$ state. Beyond the effect of the second order Stark shift, this contrast is further reduced by static electric field inhomogeneity, electric field noise and other experimental imperfections. The sensitivity to the electric field variation $\delta F$, for a fixed $\varphi_{rf}$ value, is maximal at the mid-fringe points close to the center of the interference pattern. It is clearly larger than that displayed in Fig. 2.a. 

In order to assess the improvement over the SQL, we set $\varphi_{rf}=\varphi_{rf}^0$ and we record $P(\varphi_{rf}^0, \varphi_{mw})$ as a function of $\varphi_{mw}$ for $F_0-\delta F/2$ and $F_0+\delta F/2$. Fig. 3.a presents the fringe signals together with sine fits. We extract from these fits the contrast $C$ and the relative phase $\delta\Phi=1.72$~rd of the two interference patterns, which is $21\simeq J$ times larger than the phase shift $\delta\phi$ obtained with the classical method (Fig. 2.a). We have checked that $\delta\Phi$ is proportional to $\delta F$. 

Fig. 3.b presents $\delta\Phi$ as a function of the interrogation time $\tau$, for two rf pulse durations $t_1=91$ and $t_2=184$~ns and hence two $\theta$ values (Methods). We observe that $\delta\Phi$ grows linearly with $\tau$, with a slope increasing with $\theta$. 
The experimental data are in good agreement with the predictions of Eq.~(\ref{eqn-phicat}) (dashed lines). The agreement is improved by taking into account the second order Stark effect (solid lines). Note that for $\tau\rightarrow 0$, $\delta\Phi\rightarrow \delta\Phi_0\ne0$. This is due to the finite duration of the rf pulses, during which the spin state acquires a field-dependent phase along its path on $\cal B$. 

To assess the measurement performance we only consider the phase shift $\delta\Phi_\tau =\delta\Phi-\delta\Phi_0$ accumulated during the interrogation time $\tau$.  The single-shot sensitivity is then
\begin{equation}
\sigma_F^1=\frac{1}{C}\frac{\delta F}{\delta\Phi_\tau}\ .
\end{equation}
Figure 4 compares $\sigma_F^1$ to the SQL and HL as function of $\tau$, for the $t_2$ rf pulse duration (blue points). For short interrogation times, the experimental points are well below the SQL. For larger $\tau$ values, the contrast $C$ is reduced by experimental imperfections. In fact, this reduction is in part a direct consequence of the extreme sensitivity of the measurement. Electric field noise integrated over long times blurs the interference pattern. 

The best single-shot sensitivity is $\sigma_F^1=1.2$~mV/cm for $\tau=200$ ns. The experiment repetition rate is limited by the total sequence duration, 300~$\mu$s, dominated by the atomic time of flight from preparation to detection. The sensitivity is thus 30~$\mu$V/cm/$\sqrt{\mbox{Hz}}$ corresponding to the possible detection, in 1 s, of a single electron at a 700 $\mu$m distance from the atom. This is, at least, a two orders of magnitude improvement over the sensitivity reached by NV centers \cite{Dolde2011,Dolde2014} or quantum dots \cite{Vamivakas2011,Houel2012,Arnold2014}.  Our experiment competes with the best electromechanical resonators \cite{Cleland1998,Bunch2007} or Single-Electron Transistors (SET) \cite{Yoo1997, Devoret2000}, which provide sensitivities of the order of $10^{-6}$ $e/\sqrt{\mbox{Hz}}$ at distances in the $\mu$m range, corresponding to 14~$\mu$V/cm$/\sqrt{\mbox{Hz}}$. Note, furthermore, that the experimental sequence duration could easily be reduced down to a few $\mu$s by detecting the atoms in the interaction zone. 
The sensitivity would then reach an unprecedented 3~$\mu$V/cm/$\sqrt{\mbox{Hz}}$.

We have shown that our method performs a quantum-enabled measurement of minute electric fields variations, with a non-invasive probe made of a single Rydberg atom, considerably extending the metrologic applications of these states. The measurement time is short (in the $\sim 100$ ns range) making it possible to sample tiny variations of the electric field with a MHz bandwidth. It could moreover be resolved in space, with a few micrometers resolution, using Rydberg atom excited in cold, trapped atom samples \cite{Hermann2014}.

The sensitivity could be brought much closer to the HL with an improved electrode design for a better field homogeneity and increased rf power making it possible to reach $\theta = \pi$ in spite of the second order Stark effect. An interrogation time of 200~ns would then correspond to $\sigma_F^1=170\ \mu$V/cm. For a slightly longer 1 $\mu$s interrogation time, the phase-shift of the fringes for a field increment of $200\ \mu$V/cm (that of a single electron at a 270 $\mu$m distance) reaches $\pi$, allowing in principle to distinguish two field values differing by this tiny amount with a single atomic detection (a few atoms if $C<1$).

This could lead to interesting applications in mesoscopic physics. The presence or absence of an electron in a quantum dot, realized in a 2D semiconductor or in a carbon nanotube, could be probed with a MHz bandwidth by a few atoms, far away from the mesoscopic structure. As compared to SET detectors, this method does not set tight cryogenic requirement, operates at large distances and does not require any modification of the device under test.

{\bf Acknowledgements: }  We thank A. Cottet, T. Kontos and W. Munro for fruitful discussions. We acknowledge funding by the EU under the ERC project `DECLIC' and the RIA project `RYSQ'.

{\bf Author Contributions: } A.F., E.K.D, D.G., S.H., J.M.R., M.B. and S.G. contributed to the experimental set-up. A.F. and E.K.D collected the data and analyzed the results. J.M.R., S.H., and M.B. supervised the research. S.G. led the experiment. All authors discussed the results and the manuscript.

{\bf Author Information: } The authors declare no competing financial interests.

{\bf Methods}

\paragraph{Experimental set-up}

The setup (supplementary figure 1) is made up of two parallel, horizontal disk electrodes, which create the vertical electric field $F$ aligned along the $Oz$ quantization axis. The gap between these electrodes is surrounded by four electrodes forming a ring, used to generate the rf field. The Rydberg atoms are excited stepwise by three laser beams at 780 nm, 776 nm and 1259 nm resonant with the $5S_{1/2}\rightarrow 5P_{3/2}$, $5P_{3/2}\rightarrow 5D_{5/2}$ and $5D_{5/2}\rightarrow 49F$ transitions. They cross at a 45$^\circ$ angle the horizontal atomic beam at the center of the electrode structure. The Doppler effect provides an atomic velocity selection at $v=252\pm 7$ m/s. The 780 and 776 nm cw laser beams are collinear, perpendicular to the third one.  Every 311 $\mu$s, a $0.5$ $\mu$s pulse of the 1258 nm laser excites less than one rubidium atom on average into the $49F,m=2$ state. This pulse sets the time origin $t=0$ for each sequence.  The quantization axis during the laser excitation is parallel to the 780 nm laser, and is defined by a dc field applied across the ring electrodes. This field is adiabatically switched off in 1 $\mu$s, while a $2$~V/cm field is switched on along $Oz$, which becomes the quantization axis during the measurement sequence. The atoms are then transferred in 2.7~$\mu$s into the circular state using an adiabatic rapid passage \cite{Signoles2014} in a rf field at $230$~MHz. The electric field is then ramped up in 2~$\mu$s  to $F= F_0\pm\delta F$ with $F_0=5.50527\pm 0.00021$~V/cm ($\omega(F_0)/2\pi=530.019\pm 0.020$~MHz). The state preparation sequence ends with a  0.5~$\mu$s-microwave pulse transferring  $49C$  into $50C$. This excitation, selective in the magnetic quantum number $m$, ensures that spuriously prepared elliptical states remain in the $n= 49$ manifold and do not affect the experimental signals.

The $\sigma^+$-polarized rf pulses are created by applying on two adjacent ring electrodes signals generated by 530 MHz synthesizers with finely tuned amplitudes and phases. The first pulse starts at $t= 8.5\ \mu$s.
The two microwave $\pi/2$ pulses, starting at $t=7.9$~$\mu$s and $t=23.9$~$\mu$s, are generated from the same microwave source, a frequency-multiplied X-band synthesizer. They are tuned to 51.091~GHz, on resonance with the $50C\rightarrow 51C$ transition in the $F_0$ field.

After the end of the quantum-enabled measurement, we measure the population of $\ket{J,J}$ by applying at $t=30$ $\mu$s a last $m$-selective microwave $\pi$-pulse tuned on the $50C-52C$ two-photon transition and by detecting the $52C$ level by field ionization in the detector $D$.

\paragraph{Rf pulse duration optimization} 

The rf pulses at $530$~MHz have a Rabi frequency $\Omega_{rf}/2\pi=1.6$~MHz. The 51C circular reference state $\ket R$ is the $|J',J'\rangle$ state of a $J'=25$ spin evolving on a Bloch sphere $\cal B'$ (Fig 1.a). Due to the different Stark polarizabilities in the 50 and 51 manifolds, the rf field is $11$~MHz out of resonance for the $J'$ spin ladder and barely affects it. The rf pulse results only in a small $J'$ spin precession at $\simeq 11$~MHz near the north pole of $\cal B'$. Moreover, we optimize its duration ($t_1=91$~ns or $t_2=184$~ns) so that the $J'$ spin performs exactly one or two complete rotations, returning finally to its initial $|J',J'\rangle$ state. The corresponding $\theta$ values are $\theta_1=\Omega_{rf}t_1=0.92$~rd and $\theta_2=1.86$~rd. 

\paragraph{Calibration of the electric field} 

To measure the effect induced by a variation $\delta F$ of the electric field, we alternate between an experimental sequence where we apply $F_0-\delta F/2$ and a sequence where we apply $F_0+\delta F/2$. $F_0$ is calibrated by measuring by standard spectroscopy the two transitions $\ket{49C}\rightarrow\ket{J,J}$ and $\ket{49C}\rightarrow\ket{J,J-1}$  separated by the frequency $\omega(F_0)/2\pi$. We found $\omega(F_0)/2\pi = 530.019\pm0.020$~MHz, corresponding to $F_0 = 5.50527 \pm 0.00021$~V/cm. The precision of the measurement is limited by the long term drift of the electric field (the error corresponds to the standard deviation over a few days of measurements). We also find  $\delta F = 566\pm 13$ $\mu$V/cm, corresponding to $\delta\omega/2\pi =54.8\pm1.2$~kHz.

\paragraph{Determination of the contrast $C$ of the fringes} 

The long term electric field drift affects the contrast of the interference fringes. To get their intrinsic contrast leading to the sensitivity values displayed in figure 4, we alternate sequences with $F_0+\delta F/2$ and $F_0-\delta F/2$. We then use half of the data corresponding to $F_0+\delta F/2$ to determine the slow phase drift of the interference fringes. This measured drift is used to post-process the other statistically independent half of the data, from which we deduce $C$.

\paragraph{Analytic expression of $\Phi$} 

To derive the expression of the phase $\Phi$, we consider the evolution of $\mathbf J$ in the rotating frame at frequency $\omega_{rf}$. We set the energy origin at that of the circular state $\ket{50C}=\ket{J,J}$. The first rf pulse induces a rotation $\mathcal R(\theta,0)$, preparing $ |\theta,0\rangle = \mathcal R(\theta,0) |J,J\rangle$. During the interrogation time $\tau$, $\mathbf J$ rotates along the $z$ axis of the Bloch sphere at a precession frequency $\delta\omega = \omega(F)-\omega_{rf}$, leading to the state $\ket{\theta,\phi}$, with $\phi=\delta\omega\ \tau$. Finally, a second rotation $\mathcal R (\theta,\pi+\varphi_{rf})$ brings the coherent spin state in the final state $\ket{\psi_f}$. 

The phase $\Phi$ is defined from the overlap between $\ket{\psi_f}$ and $\ket{J,J}$ :
$$\langle J,J \ket{\psi_f}= |\langle J,J \ket{\psi_f}| e^{-i\Phi} \ . $$
Using $\langle J,J \ket{\psi_f}=\langle J,J| \mathcal R (\theta,\pi+\varphi_{rf}) \ket{\theta,\phi} = \langle \theta,\varphi_{rf}\ket{\theta,\phi} $ and the expression of the scalar product of spin coherent states given in \cite{Arecchi1972} we get
$$\Phi = J\left(\phi-\varphi_{rf}  - 2 \mbox{Atan} \left[\cos\theta\tan \left(\frac {\phi-\varphi_{rf} } 2\right) \right]\right) $$

In the classical method, the atom is initially prepared in $\ket {J,J}$, and the probability to find it in $\ket{J,J}$ at the end of the sequence is $$ P(\varphi_{rf}) = |\langle \theta,\varphi_{rf}\ket{\theta,\phi}|^2 = \exp(-J\sin^2\theta(\phi-\varphi_{rf})^2/2)\ .$$

In order to measure $\Phi$, we prepare, with a first mw $\pi/2$ pulse, a quantum superposition of $\ket{J,J}$ and of the reference state $\ket R$. We then apply a second $\pi/2$ pulse resonant with the $\ket{J,J} \rightarrow \ket R$ transition after the rf pulses. The probability to find the atom in $\ket{J,J}$ is then given by :
$$ P(\varphi_{rf},\varphi_{mw}) = \frac 1 4+ \frac 1 4P(\varphi_{rf})+ \frac 1 2\sqrt{P(\varphi_{rf})}\cos(\Phi-\varphi_{mw})\ ,$$
where $\varphi_{mw}$ is the relative phase between the mw pulses, and $\Phi$ implicitly depends on $\varphi_{rf}$.

The probability to find the atom in $\ket{J,J}$ therefore oscillates with $\Phi$, with an amplitude proportional to $\sqrt{P(\varphi_{rf})}$. For small values of $\delta\omega$, this amplitude is maximum for $\varphi_{rf}\approx 0$ and then :
$$\Phi \approx \Phi_0 + J \,(1 - \cos\theta)\,\left(\frac{\partial\omega}{\partial F}  \right) \delta F \,\,  \tau\ \ $$ where $\Phi_0$ is the phase accumulated for $\delta\omega = 0$.

\paragraph{Single atom sensitivity} 

The single-shot sensitivity is given by $\sigma_F^1=(\partial P(\varphi_{rf},\varphi_{mw})/\partial F)^{-1} \sigma_P$ where $\sigma_P$ is the dispersion of an atomic state detection. It can be rewritten as  
\begin{equation}
\sigma_F^1=\left(\frac{\partial P}{\partial \Phi}\right)^{-1}\left( \frac{\partial \Phi}{\partial F}\right)^{-1}\sigma_P
\end{equation}

The optimum strategy to measure the electric field is to set the phase $\varphi_{rf}$ that maximizes the contrast of the fringes, and to set $\varphi_{mw}$ so that $P=1/2$ (mid-fringe setting). Therefore $\sigma_P= \sqrt{P(1-P)}=1/2$ is maximal.

In the ideal case, ${\partial P}/{\partial \Phi} = 1/2$, $\partial \Phi/\partial F =\tau J  (1-\cos\theta) (\partial \omega/\partial F)$, leading to a theoretical single-shot sensitivity :
\begin{equation}
\sigma_F^{1,th}=\frac{1}{\tau J (1-\cos\theta)}  \left(\frac{\partial \omega}{\partial F}\right)^{-1}
\end{equation}

We calculate the experimental sensitivity corresponding to the interrogation time $\tau$ by considering only the differential phase $\delta\Phi_\tau$ accumulated between the two rf pulses, and writing $\partial \Phi/\partial F\approx \delta\Phi_\tau/\delta F$. We also take into account that ${\partial P}/{\partial \Phi} = C/2$ is reduced by the finite contrast $C$ of the interference fringes. Finally, $\sigma_F^{1,exp}= 1/C\cdot \delta F/\delta\Phi_\tau$.

\clearpage

\begin{figure}

   \includegraphics[width=\linewidth]{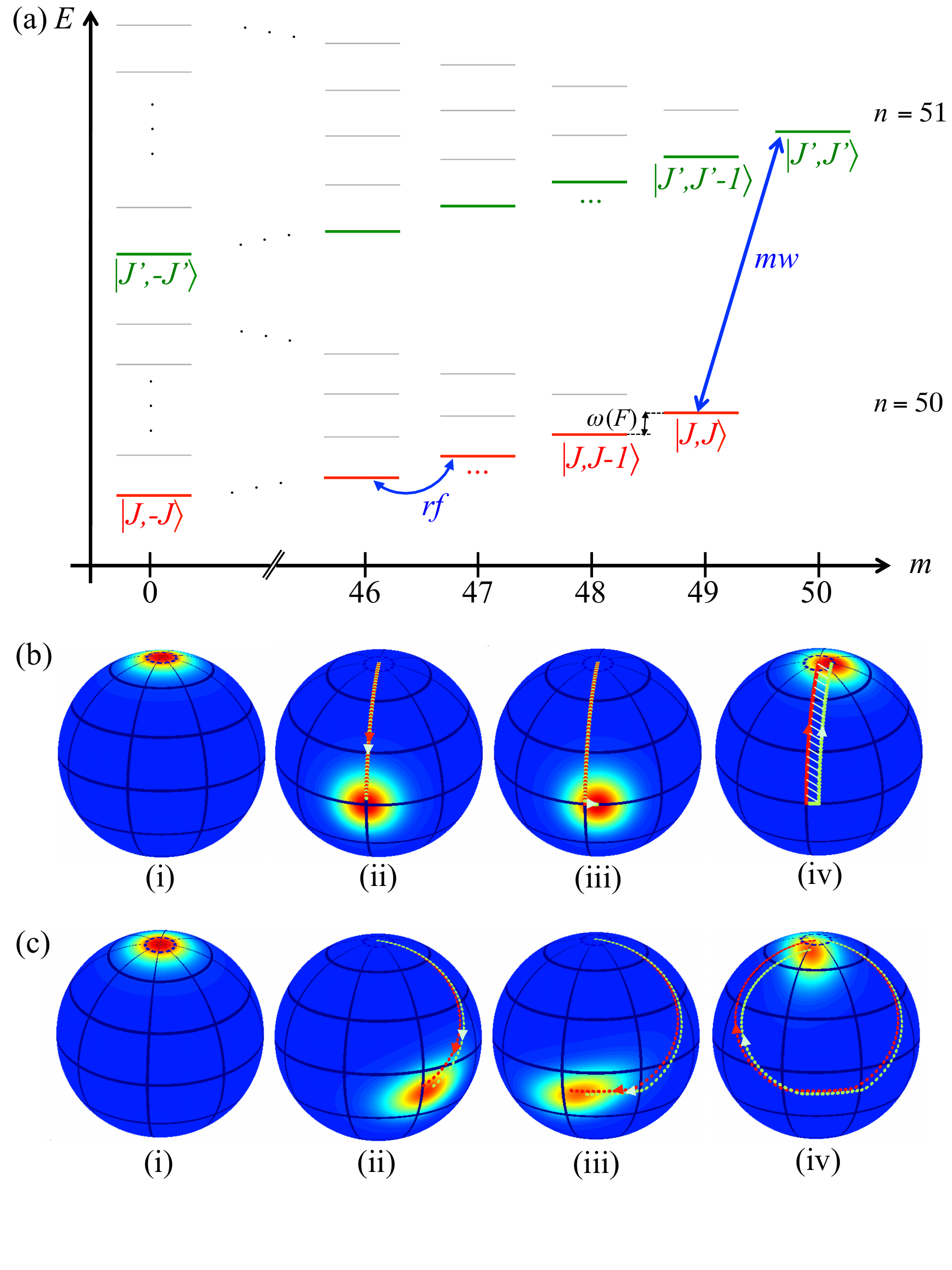}
\caption{{\bf Atomic levels and measurement sequence.} (a) Energies of the Stark levels in the 50 and 51 manifold sorted by their magnetic quantum number $m$ for $m\ge0$ (not to scale). The $J$ spin states in the 50 manifold are depicted by the thick red lines. The reference state is the circular state $\ket{J',J'}$ in the 51 manifold belonging to the $J'$ state ladder (green lines). (b) Evolution of the $J$ spin in the Ramsey sequence (Stark effect to first order in $F$). Successive plots of the spin state  $Q$-function in the rotating frame for $F=F_0+\Delta F$. (i) Initial $\ket{J,J}$ circular state. (ii) State after the first rf pulse inducing a $\theta=\pi/2$ rotation. (iii) State after the interrogation time $\tau$, before the second rf pulse. (iv) Final state after the second rf pulse with $\varphi_{rf}=0$. The green dotted line shows the trajectory on the Bloch sphere $\cal B$. The red line corresponds to the spin trajectory for $F=F_0$. The classical Ramsey scheme measures $\Delta F$ through the final position of the spin in (iv), and is therefore limited by the quantum fluctuations of the SCS. In the quantum enabled scheme, we deduce $\Delta F$ from the global phase accumulated during the complete evolution, proportional to the dashed area.
(c) Simulation of a realistic sequence for $\tau = 56$~ns and $\Delta F= 1.7$ mV/cm, taking into account the second-order Stark effect and the finite duration of the rf pulses (184 ns).  The phase $\varphi_{rf}$ is chosen so that the trajectory is closed for $F=F_0$. The $Q$-function and the green line corresponds to $F=F_0+\Delta F$. The red line corresponds to the spin trajectory for $F=F_0$. The value of $\Delta F$ is deduced from the difference of the global phases accumulated along the red and green trajectories.
}
\label{fig-1}
\end{figure}

\begin{figure}
 \includegraphics[width=\linewidth]{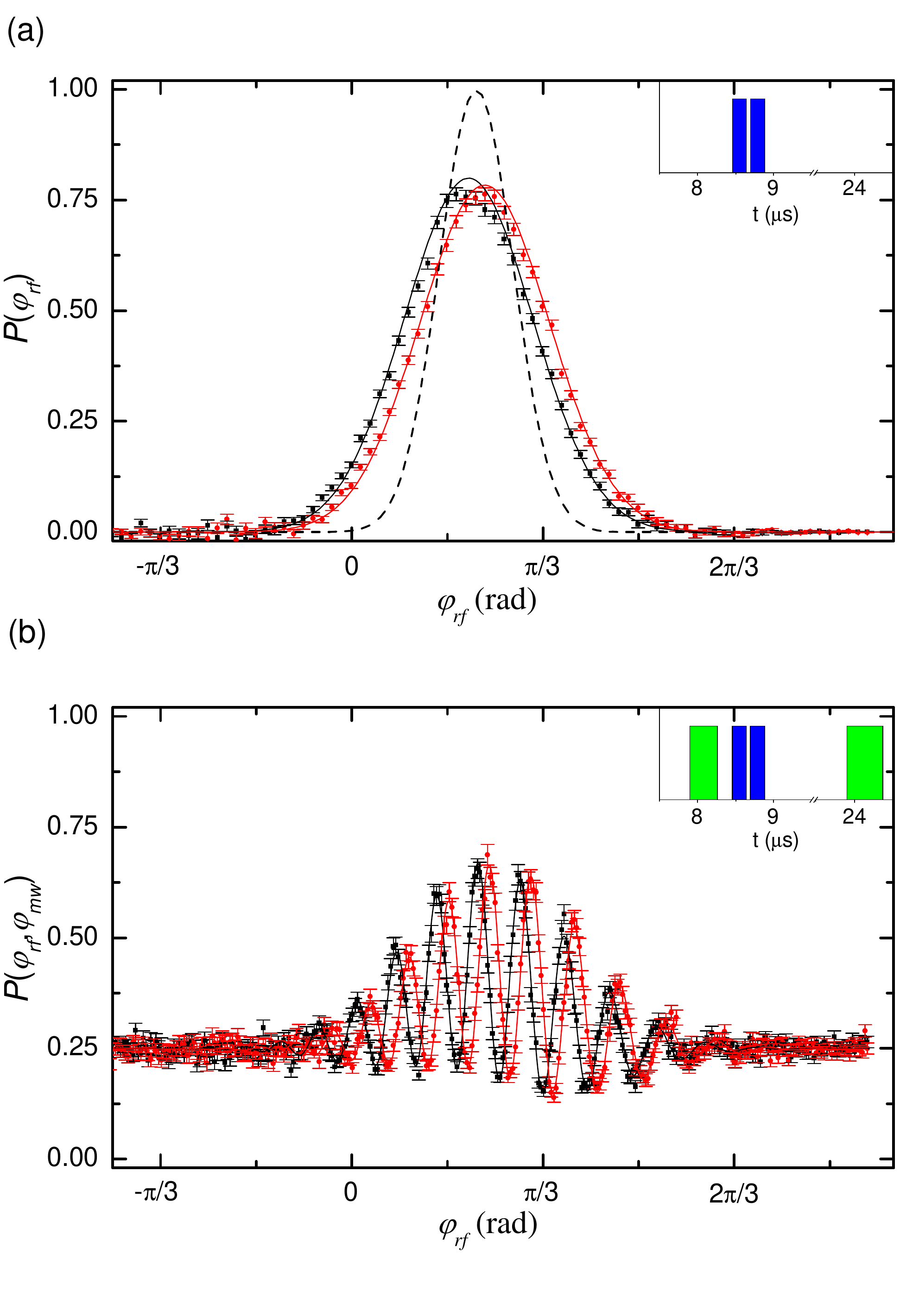}
  \caption{{\bf Classical and quantum-enabled field measurements.} (a) Simple Ramsey measurement. Probability $P(\varphi_{rf})$ for ending up in $|J,J\rangle$. The rf pulses (blue bars in the timing inset), separated by the interrogation time $\tau=56$~ns, have the duration $t_2=184$~ns and correspond to $\Omega_{rf} t_2=1.86$~rd (Methods). The red and black experimental points correspond respectively to the field values $F_0+\delta F/2$ and $F_0-\delta F/2$, with $\delta F=566$~$\mu$V/cm. The two signals are shifted by $\delta\phi = 82$ mrad. The reference phase $\varphi_{rf}^0$ is determined as the average of the centers of these curves. The error bars reflect the statistical uncertainties over 3100 realizations of the experiment. The solid lines result from numerical simulations of the full experiment. The dotted line is a Gaussian with a width determined by the SQL. (b) Results of 950 realizations of the same experiment including the microwave pulses. The timing is in the inset (green : mw pulses, blue: rf pulses). The data points are experimental, with statistical error bars, with solid lines to guide the eye. The signals corresponding to $F_0+\delta F/2$ and $F_0-\delta F/2$ are shifted by about the same $\delta\phi$ as in (a). The interference fringes spacing being much smaller than the width of the Gaussian in (a), the quantum-enabled measurement is much more sensitive to variations of the electric field. The background probability for $\varphi_{rf}$ far from $\varphi_{rf}^0$ is 1/4. When the spin does not return close to the initial state, the $n=50$ manifold does not contribute to the signal. However, half of the 50\% population stored in $n=51$ returns in $n=50$ after the final mw pulse.}
\label{fig-2}
\end{figure}

\begin{figure}
\includegraphics[width=\linewidth]{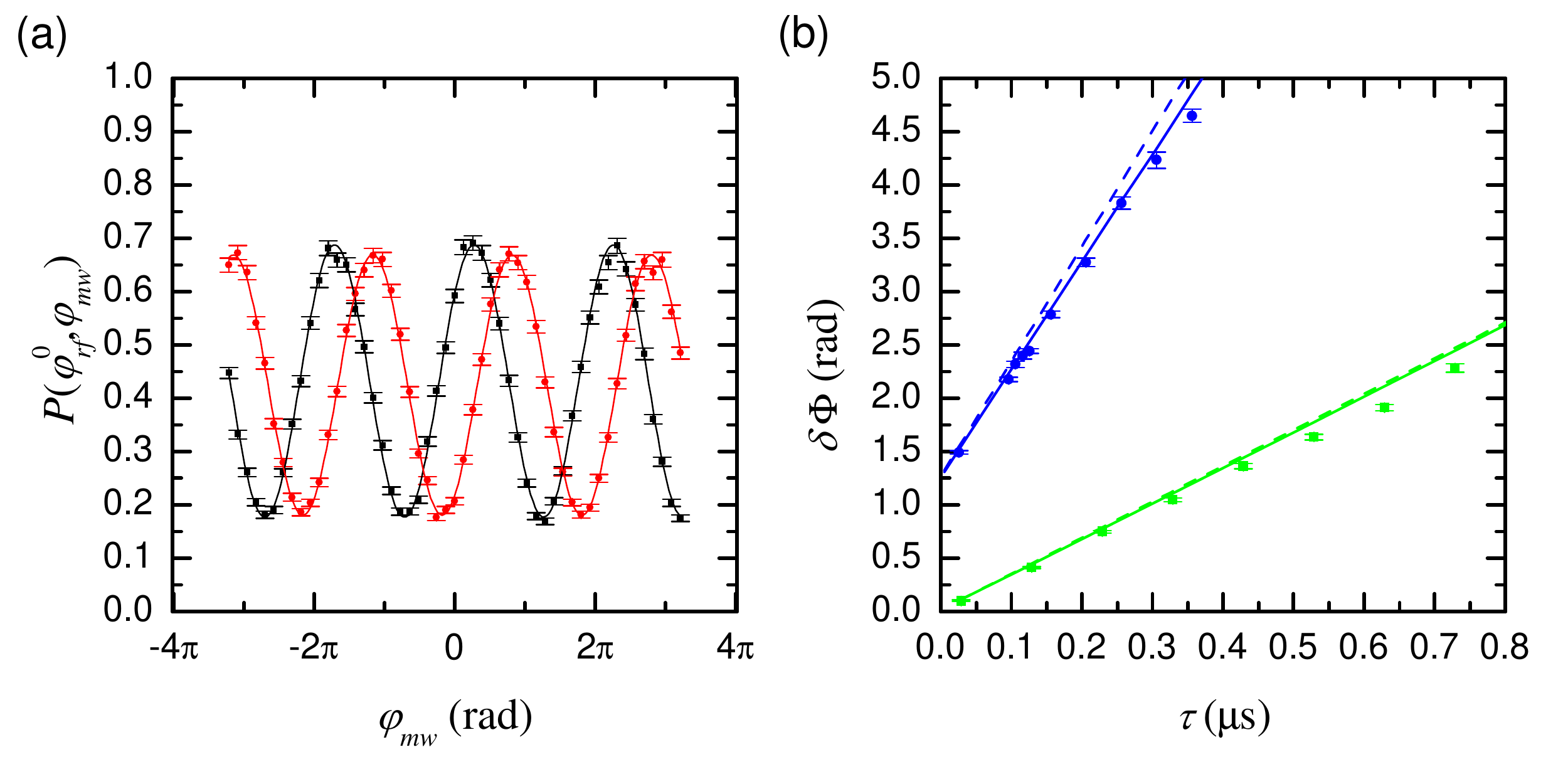}

\caption{{\bf Microwave Ramsey fringes} (a) Probability $P(\varphi_{rf}^0,\varphi_{mw})$ for being in $|J,J\rangle$ at the end of the complete sequence [experimental conditions are the same as for Fig. 2.(b)], as a function of the relative phase $\varphi_{mw}$ of the mw pulses. The red and black dots correspond to experiments at $F_0+\delta F/2$ and $F_0-\delta F/2$ respectively (error bars are statistical over 3100 realizations). The solid lines are sine fits, providing the relative phase $\delta\Phi$ and the contrast $C$ of the interference signals. (b) Phase increment $\delta\Phi$ corresponding to $\delta F$ as a function of the interrogation time $\tau$. The green and blue points (error bars resulting from the fringe fits) correspond to the rf pulse durations $t_1$ and $t_2$ respectively. The slope of the dashed lines correspond to the ideal model with no quadratic Stark effect. The solid lines are predictions of a numerical model including the second order Stark effect, which distorts the trajectory on $\cal B$ and slightly reduces the value $\theta$ w.r.t. the ideal value $\Omega_{rf} t_i$ ($i=1,2$). 
     }
\label{fig-3}
\end{figure}

\begin{figure}
\includegraphics[width=\linewidth]{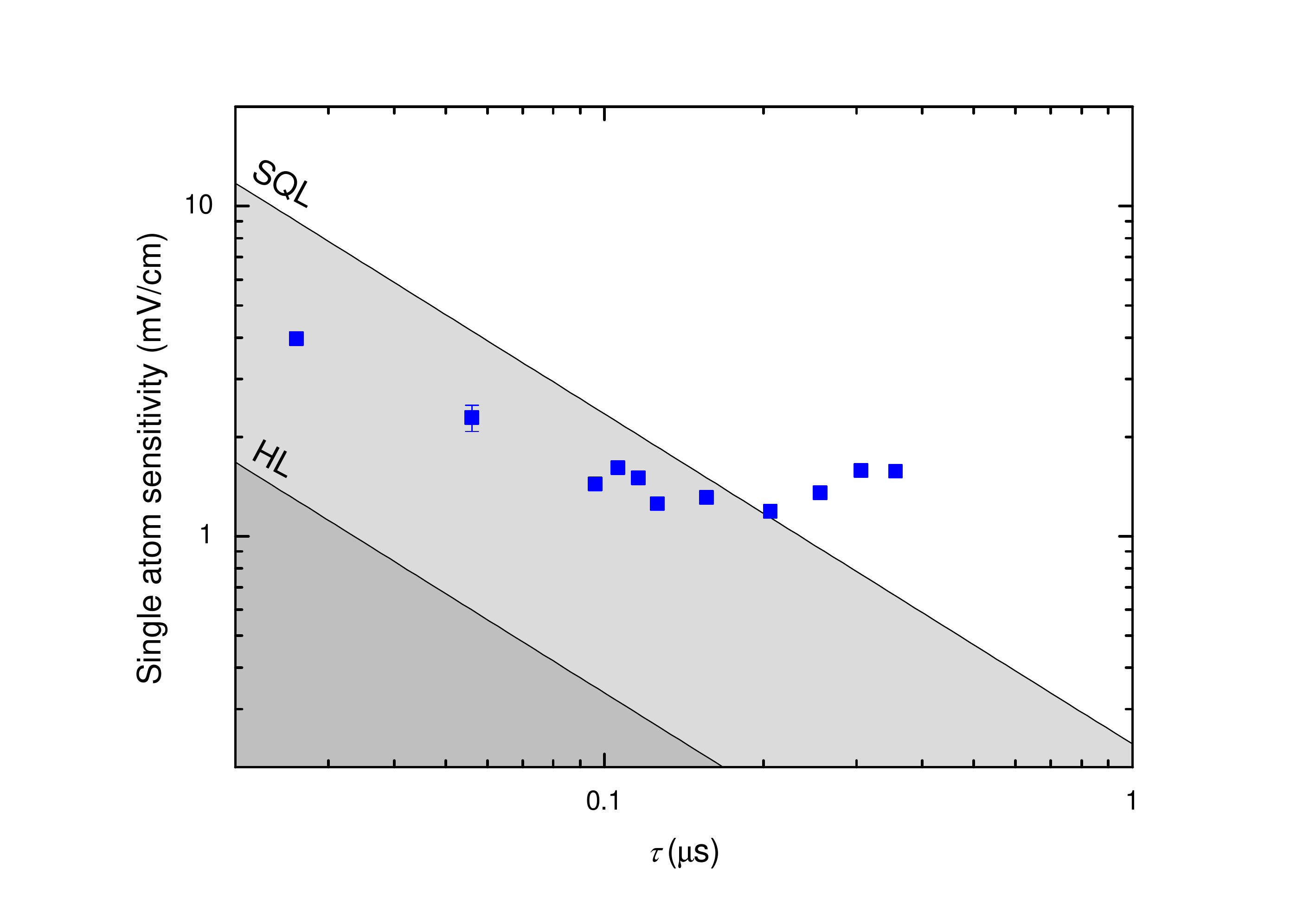}
  \caption{{\bf Comparison of the single-shot sensitivity with the SQL and HL.} The blue points show a logarithmic plot of $\sigma_F^1$ as a function of the interrogation time $\tau$. The error bar on the point at $\tau=56$~ns reflects the statistical variance of three experiments performed in the same conditions. The SQL and HL  are depicted by black lines. The dark gray area is forbidden. Quantum-enabled measurements lie in the light-gray area. This measurement outperforms the SQL for $\tau<200$~ns, by up to a factor 2 for the shortest time.
  }
\label{fig-4}
\end{figure}
\setcounter{figure}{0}
\renewcommand{\figurename}{EXTENDED FIGURE}

\begin{figure}
 \includegraphics[width=\linewidth]{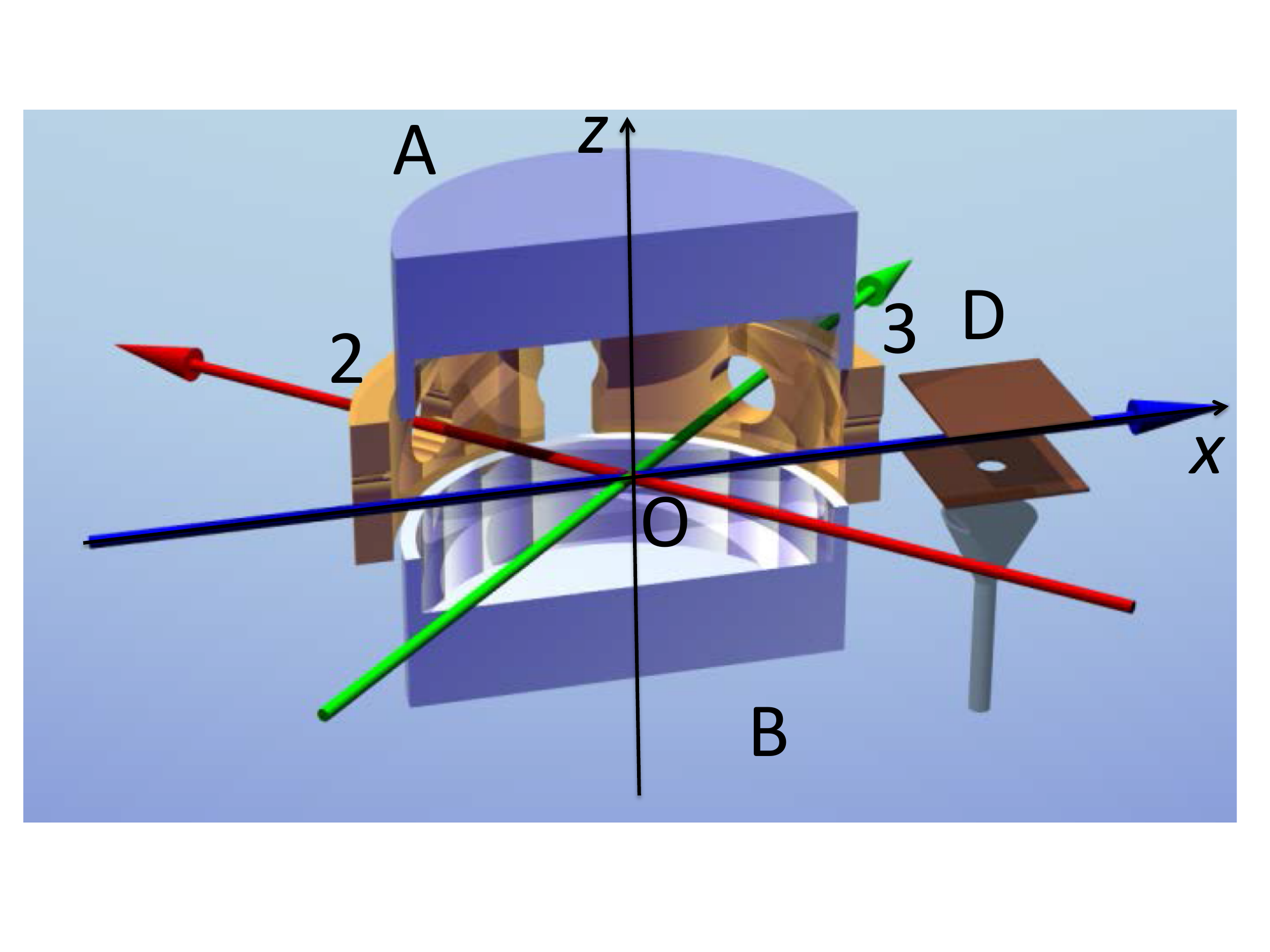}
  \caption{{\bf Schematic of the experiment} 
  The atoms are produced by excitation of a thermal Rubidium beam (blue arrow) propagating along axis $Ox$. Two horizontal electrodes $A$ and $B$ (represented here cut by a vertical plane) produce the directing electric field $\mathbf{F}$ along $Oz$. The gap between $A$ and $B$ is surrounded by four independent electrodes (1, 2, 3 and 4), on which we apply RF signals to produce $\sigma_+$ fields with tunable phase and amplitude. Electrodes 1 and 4, not represented, are the mirror images of electrode 2 and 3 (in yellow) with respect to the $xOz$ plane.
 The laser excitation to the Rydberg states is performed using three laser beams that intersect in the center $O$ of the cavity. The 780 nm and 776 nm laser beams are collinear (red), the 1258 nm laser is sent perpendicular to the other beams (green). 
Once the atoms have left the electrode structure, they enter the field-ionization detector $D$.}
\end{figure}

\end{document}